\documentclass[namedreferences]{SolarPhysics}
\usepackage{natbib}                     	% For citations: redefine \cite commands
\usepackage[optionalrh]{spr-sola-addons} 	% For Solar Physics
\usepackage{solaheader} 			% A header, appropriate for the pre-print servers
\usepackage{graphicx}
\usepackage{rotating}
\usepackage{courier}  				% Change the \texttt command to courier style
\usepackage{amssymb}                    	% useful mathematical symbols
\usepackage{url}                         	% For breaking URLs easily trough lines
\newcommand{\etal}{et al. }
\newcommand{\be}{\begin{equation}}
\newcommand{\ee}{\end{equation}}
\newcommand{\beq}{\begin{eqnarray}}
\newcommand{\eeq}{\end{eqnarray}}
\newcommand{\adv}{    {\it Adv. Space Res.}}

\newcommand{\aap}{    {\it Astron. Astrophys.}}
\newcommand{\araa}{   {\it Ann. Rev. Astron. Astrophys.}}

\newcommand{\aapr}{   {\it Astron. Astrophys. Rev.}}

\newcommand{\apj}{    {\it Astrophys. J.}}
\newcommand{\apjl}{   {\it Astrophys. J. Lett.}}

\newcommand{\jastp}{  {\it J. Atmos. Sol. Terr. Phys.}}
\newcommand{\jgr}{    {\it J. Geophys. Res.}}

\newcommand{\pasj}{   {\it Pub. Astron. Soc. Japan}}

\newcommand{\solphys}{{\it Solar Phys.}}

\newcommand{\ssr}{    {\it Space Sci. Rev.}}
\newcommand{\aplett}{ {\it Astrophys. Lett.}}
\newcommand{\todash}{\,--\,}
\newcommand{\solareceiveddate}{08 December 2009} % Remove the leading % and fill in the values when known; solaheader must be used
\newcommand{\solaaccepteddate}{28 September 2010}
\newcommand{\doi}{10.1007/s11207-010-9648-7}
\begin{document}
\begin{article}
\begin{opening}
\title{Radio Observations of the 20 January 2005 X--class flare}
%%%%%%%%%%%%%%%%%%%%%%%%%%%%%%%%%%%%%%%%%%%%%%%%%%%
%% Authors Names
%
\author{C. \surname{Bouratzis}$^{1}$\sep
	P. \surname{Preka-Papadema}$^{1}$\sep
	A. \surname{Hillaris}$^{1}$\sep
	P. \surname{Tsitsipis}$^{3}$\sep
	A. \surname{Kontogeorgos}$^{3}$\sep
	V. G. \surname{Kurt}$^{2}$\sep
	X. \surname{Moussas}$^{1}$}
%
%Y ushkov, B. Yu.;Skobeltsyn Institute of Nuclear Physics, Lomonosov Moscow State University
% 	clef@srd.sinp.msu.ru
%-------------------------------------------------
%% Runningheads
\runningauthor{C. Bouratzis \emph{\etal}}
\runningtitle{Radio Observations of the 20 January 2005 X--class flare}
%%%%%%%%%%%%%%%%%%%%%%%%%%%%%%%%%%%%%%%%%%%%%%%%%%%
%% Affilations
\institute{$^{1}$ Section of Astrophysics, Astronomy and Mechanics, Department of Physics,
		University of Athens, Zografos (Athens) , GR-15783, Greece
		email: \url{ppreka@phys.uoa.gr} email: \url{kbouratz@phys.uoa.gr} 
		email: \url{ahilaris@phys.uoa.gr} email: \url{xmoussas@phys.uoa.gr}\\
           $^{2}$ Skobeltsyn Institute of Nuclear Physics, Lomonosov Moscow State University, Moscow, 119992, Russia 
		email: \url{vgk@srd.sinp.msu.ru}\\
           $^{3}$ Department of Electronics, Technological Educational Institute of Lamia, Lamia , GR-35100, Greece
		email: \url{tsitsipis@teilam.gr} email: \url{akontog@teilam.gr} \\}
%-------------------------------------------------
\begin{abstract}
We present a multi-frequency and multi-instrument study of the 20 January 2005
event. We focus mainly on the complex radio signatures and their association
with the active phenomena taking place: flares, CMEs, particle acceleration and magnetic 
restructuring. As a variety of energetic particle accelerators and sources of
radio bursts are present, in the flare--ejecta combination, we investigate their relative
importance in the progress of this event. The dynamic spectra of \emph{Artemis--IV--Wind$/$Waves--Hiras} 
with 2000 MHz-20 kHz frequency coverage, were used to track the evolution of
the event from the low corona to the interplanetary space; these were supplemented with SXR, HXR
and $\gamma$-ray recordings. The observations were compared with the expected radio
signatures and energetic-particle populations envisaged by the {\em{Standard Flare--CME model}} and
the {\em{reconnection outflow termination shock}} model.
A proper combination of these mechanisms seems to provide an
adequate model for the interpretation of the observational data.

\end{abstract}
%-------------------------------------------------
%% Keywords
\end{opening}
%-------------------------------------------------
\section{Introduction}\label{Introduction}

Solar flares and coronal mass ejections (CMEs) are the most energetic phenomena
on the Sun. They can be envisaged as two different aspects of a
common magnetic energy  instability and release \citep[e.g.][and references therein]{Pick06},
described by  the standard CME-flare model which combines
the kinematics of the  energy release of ejecta and flares: 
As the rising CME (or erupting filament or rising plasmoid) stretches
and deforms the coronal magnetic-field, a vertical current sheet is formed,
where reconnection explosively releases the free magnetic energy, stored
in the corona \citep[e.g.][]{Forbes00, Forbes03, Priest02}; furthermore, 
CME acceleration profile and flare energy release are seen to evolve in a
synchronized manner \citep{Temmer10}. The energy, thus liberated,
is divided over plasma heating, particle acceleration, 
kinetic energy of the eruption, and MHD shock waves. 

The development, in the low corona, of large flare$/$CME events
coincides with an extended opening of the magnetic field, accompanied by
energetic-particle acceleration and injection into interplanetary space and shocks.
These are detectable, at metric and longer waves, by their radio signatures
\citep[e.g. review by][]{Pick08}.

The Type III burst radio emission is produced by non-thermal electrons streaming along
coronal magnetic lines. In open field lines, they may escape into interplanetary
space and may be detected {\em{in situ}} \citep[e.g.][]{Fainberg70, Lin73, Kurt77, Klein08}. 
In closed magnetic structures, on the other hand, they eventually turn Sunwards
resulting in inverted U- or J-shaped bursts on the dynamic spectra (U or J bursts). 
Metric bursts of the Type III family generally consist of groups of individual bursts
\citep[e.g. ][ and references within]{Goldman83}; on the reports they are marked 
as III G for fewer than 10 occurences and III GG for more.

The Type II bursts trace the propagation of MHD shocks in the corona and 
interplanetary space. It is, in general, accepted that Type II bursts at decametre
and longer wavelengths are driven by CMEs, bow or flank,
\citep [e.g. ][for a review]{Vrsnak08}. The Type II bursts at metre wavelengths, however, are 
interpreted either as a flare blast wave \citep{Vrsnak01} or a CME driven 
shock \citep{Kahler84,Maia00,Classen}. A metric stationary Type II burst
(a Type II with a very slow drift on the dynamic spectrum)
has been identified, for the first time, by \citet{Aurass02}; 
it was interpreted as the radio signature of a fast-mode termination shocks by
\citet{Aurass06}. The proposed formation process  included
a pair of counterstreaming outflow jets 
moving upwards and downwards from the reconnection site, 
encountering  the rear of the CME and the top of the post flare
loop, respectively, resulting in termination shocks \citep{Mann06, Mann09, Warmuth09}. 

The continua during periods of activity, represent the radiation of energetic electrons trapped
within magnetic structures and plasmoids, and they are known under the names of Type IV bursts and
``Flare Continua'' \citep{Robinson78B,Robinson85}. The stationary Type IV (IV mB) bursts emanate
from magnetic structures usually located above active regions; they often
exhibit significant fine structure. The moving Type IV bursts \citep{Robinson78A}
(IV mA) are emitted from sources of metre wave continuum moving outwards at velocities of the
order of 100\todash1000 km$s^{-1}$; their spectrum is often featureless and sometimes lasts more than ten minutes.
A number of these are believed to originate within the dense substructures 
(such as erupting prominences) within the CMEs \citep{KleinMouradian02,Bastian01}.
Others appear following Type II bursts and are possibly caused by energetic
electrons produced in the wake of the Type II shock. \citet{Robinson85} calls them
flare continua II (FCII) based on the close temporal association with the Type II
and classifies them in the stationary Type IV (IV mB) family with frequency drift.

Broad-band ({{$\Delta$}f/f $\approx$ 1}) and narrow-band ({{$\Delta$}f/f $<$ 0.1}) pulsations
are, at times, considered  part of the Type IV family being interpreted as MHD oscillations in coronal
loops modulating  the continuum. Alternative interpretations include limit cycles of nonlinear wave--particle
interactions in coronal loops or quasi-periodic particle
acceleration episodes during the magnetic reconnection
in a large-scale current sheet. On the dynamic spectra, the pulsations
exhibit quasi-periodic behavior ($\approx$ 0.1--10 Hz) and durations from
seconds to minutes; some gradually drift to lower frequency in the course of the
event \citep[e.g. reviews by][]{Benz03,Nindos07}.
%--------------------------------------------------------------------------------------

Another type of radio burst which has been associated with magnetic reconnection
in solar flares is the short ($\le$ 0.1 second), narrow-band (some MHz) emissions known as
``spikes''. They appear in  broad-band clusters lasting seconds to about a minute.
Metric spikes at 250$-$500 MHz are quite common near the starting frequency of metric Type III
bursts and are thus linked with the acceleration of electron beams \citep{Benz03,Kliem03}.
They are considered as the signatures of accelerated particles
at a highly fragmented,  primary energy release site \citep{Nindos07}.
At the same time they pinpoint the location of the reconnection and acceleration site.
%--------------------------------------------------------------------------------------

A small number of solar active phenomena has been classified as ``extreme events'';
in these events, characteristics such as flare intensity, CME speed and solar-energetic
particles (SEP) flux are orders of magnitude larger than the rest \citep{Crosby09}.
Solar-energetic proton events may result in short duration
sharp increase in the count rate of ground based cosmic ray
detectors and are dubbed ground level enhancements (GLEs).

The extreme event of 20 January 2005 in AR 720 (N14$^o$ W61$^o$),
included a fast CME, an X7.1$/$2B (6:36\todash7:26 UT) flare, a white-light flare, 
hard X-rays, $\gamma$-rays (up to 300 MeV), and radio bursts in microwave 
to decametric frequency. It also initiated SEP and GLE, indicating
a prompt proton acceleration \citep{Kuznetsov06,Kuznetsov08,Grechnev08,Simnett07}
with the hardest energy spectrum of solar cycle 23 \citep{Bazilevskaya09}.
The GLE, in particular, was characterized as the most intense in half a century and among the
most intense in observation history \citep{Plainaki07,Belov05B},
second only to the GLE of 23 February 1956 \citep{Vashenyuk05a}.

This report on the 20 January 2005 extreme event is structured as follows:
In section \ref{Data} we describe the instrumentation and data set used in our study;
this is supplemented with an algorithm for the detection and statistical analysis of
quasi-linear structures embedded in complex dynamic spectra in the appendix. The
observational results are presented in Section \ref{Event}; they are discussed
in Section \ref{SandD}  and compared with
models of magnetic reconnection and particle acceleration within the framework of the
CME-induced  reconnection. Conclusions are presented in Section \ref{Con}.

\section{Instrumentation and Data Processing} \label{Observations} \label{Data}

\subsection{Instruments and Data Sets}

The Appareil de Routine pour le Traitement et l'
Enregistrement Magnetique de l' Information Spectral (Artemis-IV) solar
radio$-$spectrograph(\url{http//web.cc.uoa.gr/artemis/})
operating at Thermopylae Greece ($38^o 49'$ N, $22^o 41'$ E),
since 1996 \citep{Caroubalos01,Kontogeorg06}
consists of a 7-metre parabolic antenna covering the metric range and a dipole
antenna covering the decametric range. Two
receivers operate in parallel, there are a sweep frequency analyzer (ASG), 
covering the 650-20 MHz range in 630 data channels at ten samples~$sec^{-1}$ and a
high sensitivity multi-channel acousto-optical analyzer (SAO), which covers
the 270-450 MHz range in 128 channels at 100 samples~$sec^{-1}$. 
The broad-band, medium time resolution recordings of the ASG are used for the detection and analysis of
radio emission from the base of the corona to two $R_{\odot}$, while the narrow-band, high time
resolution SAO recordings are mostly used in the analysis of the fine temporal and spectral structures.

In this work we used data from the Hiraiso Radiospectrograph (Hiras) \citep[25\todash2500 MHz, ][]{Kondo95}
to complement the Artemis-IV range in the high-frequency domain.

The {\em{Wind$/$Waves}} experiment \citep{Bougeret95} includes the RAD2 receiver which
covers the critical frequency range from 13.825 MHz ($R_{\odot}$)
to 1.075 MHz (20 $R_{\odot}$). It is complemented by the RAD1 receiver
(frequency range 1040-20 kHz), which allows us to track the evolution of the radio sources to 1 AU.

The dynamic spectra were supplemented with recordings in the 2-35 GHz range of the
Nobeyama Radio Polarimetre (NoRP) \citep*{Nakajima85}. The event was also 
observed on HXR and $\gamma$-rays in the 0.03-300 MeV range from the
Solar Neutron and Gamma (SONG)$/$CORONAS-F \citep*{Kuznetsov04}; we used 
RHESSI data from \citep*{Grechnev08}. The SXR data were obtained from 
GOES (\url{http//www.sel.noaa.gov/ftpmenu/indices}) and the kinematics of
the CME came from \citet{Gopalswamy05}.

The times of energetic-particle release from the low corona
were determined from the reports of \citet{Saiz05, Grechnev08, Simnett07, Kurt10a, Kurt10}.

Thus, with {\em{Wind$/$Waves}} observing the Interplanetary Space and HiRAS with  
Artemis-IV the high-frequency counterpart of the event
(corona and lower corona) we obtain a 2000 MHz$-$20 kHz frequency coverage.
This, combined with the SXR, HXR and $\gamma$-ray recordings, makes a
multi-frequency and multi-instrument study of all different aspects of the 
active phenomena possible, the injection of energetic particles included, comprising the 20 January 2005 event.
%-------------------------------------------
% a landscape figure which scales and rotates the file f01.eps
\begin{sidewaysfigure}
\centering
\resizebox{\textwidth}{!}{\includegraphics{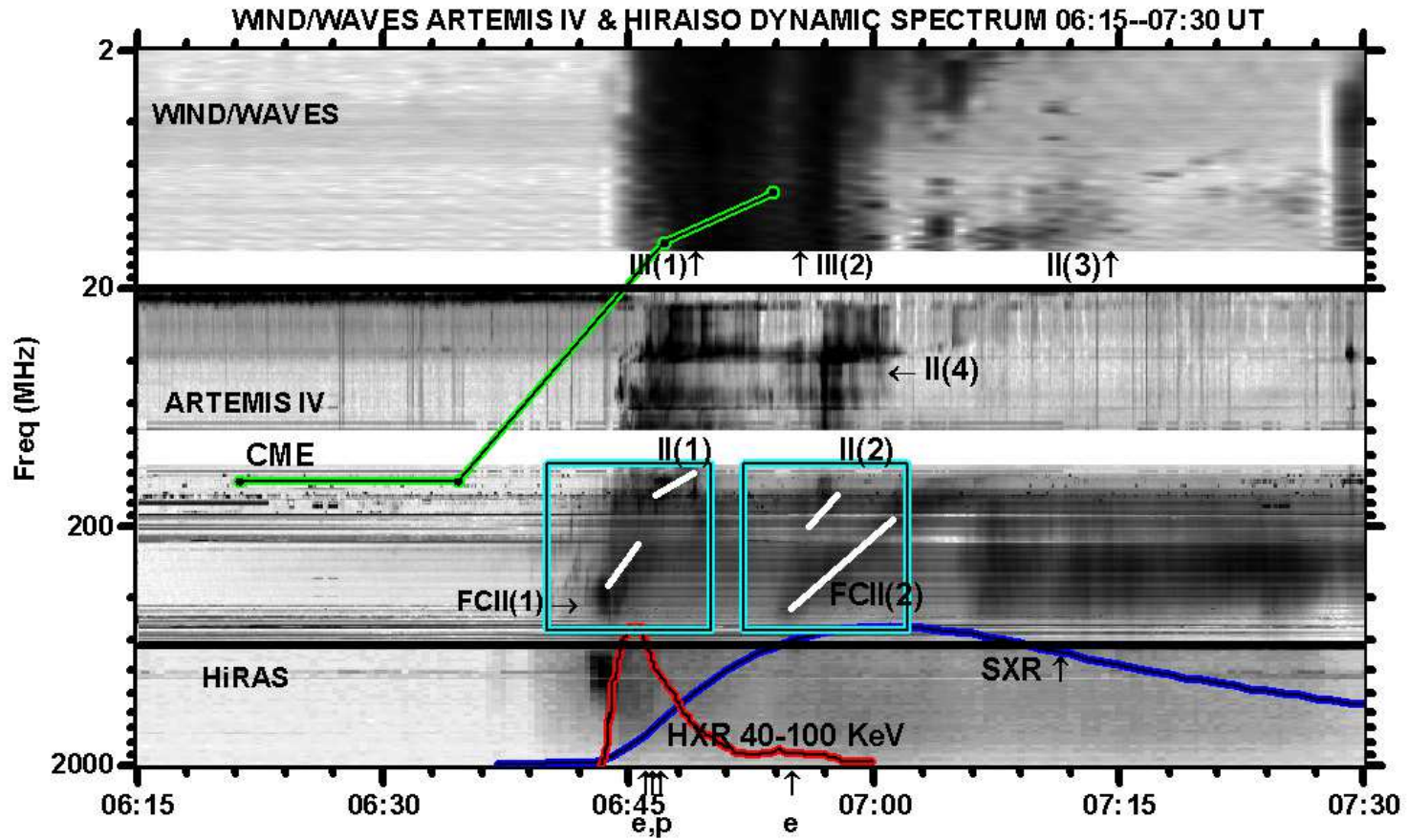}}
\caption{Combined HiRAS, Artemis-IV (ASG), and  {\em{Wind$/$Waves}} dynamic spectrum of the 
event of 20 January 2005, 06:15-07:30 UT and 2000-2 MHz; the CME 
trajectory using the Newkirk model (see text) for the
height to frequency conversion (green); the GOES SXR flux (blue); the SONG 40-100 KeV channel (red);
the two Type II$/$FCII combinations (II(1)$/$FCII(1) and II(2)$/$FCII(2)) are in cyan frames; 
details are presented in Figure \ref{fig01B}. The electron and proton release times
as reported by \citet{Grechnev08,Simnett07} and \citet{Saiz05} (Table \ref{T}) are 
annotated with arrows under the plot.}
\label{fig01}
\end{sidewaysfigure}
%------------------------------------------------
\begin{figure}
\resizebox{\textwidth}{!}{\includegraphics{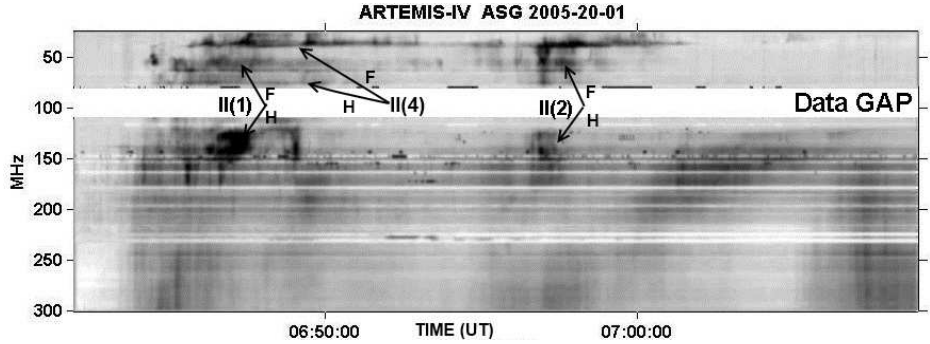}}
\caption{Artemis-IV$/$ASG dynamic spectrum (linear frequency scale 300-40 MHz)
of fundamental-harmonic bands of the two metric (II(1) and II(2)) and the decametric (II(4)) 
shocks }
\label{Lin}
\end{figure}
%-------------------------------------------
\begin{table}
\centering
\caption{ Overview of the Event 20 January 2005 and associated activity.}
\label{T}
\begin{tabular}{lllll}
\hline
  Event        		& UT 			& Characteristics  		& Remarks \\
\hline
CME Lift Off		& 06:33 		& 				&\citet{Gopalswamy05} \\
Start			& 06:36 		& 3.75-35.0 GHz		& NoRP \\
SXR Start    		& 06:36 		& AR 720: N14$^o$ W61$^o$	& GOES 12 \\
IV Start		& 06:36 		& 2000-100 MHz			& Artemis-IV, Hiras \\	
\hline
 Narrow-band		& 06:39 		& Turnover $\approx$200 MHz 	& Artemis-IV \\
 J and U		&   			& Drifting to $\approx$115 MHz	& SAO \\
 H$\alpha$ Start  	& 06:41 		& Two ribbon, AR 720		& LEAR \\
HXR increase  		& 06:42			& $\approx$100 keV		& \citep{Masson09}\\
 Spikes Start  		& 06:42:40 		& 450-270 MHz			& SAO \\
Continuum Patch		& 06:43			& 460-350(MHz)			& Artemis-IV \\
 (F$-$H)	 	& to 06:43:45      	& 910-690(MHz)			& and Hiras\\
FCII(1) Start 		& 06:43			& 650 MHz, 	    		& Artemis-IV \\
        		&  			&  340 Km~sec$^{-1}$		&  \\
$\gamma$-ray increase 	& 06:43-06:44		& $\approx$ 780 KeV-10 MeV	& \citep{Kurt10}\\
Microwave Peak		& 06:45 		& 35 GHz			& NoRP \\
$\gamma$-ray increase	& 06:45-06:46		& $\approx$ 60-100 MeV		& SONG$/$CORONAS-F\\
			&			&				& \citep{Kurt10}\\
 II(4) Start  		& 06:45:16 		& 41 (F)-82 (H) MHz		& Artemis-IV \\
   HXR peak    		& 06:45:30 		& $\approx$100 keV		& RHESSI \\
  III(1) GG      	& 06:45:39		&				& Artemis-IV, Hiras \\
			&       		& 				& and {\em{Wind$/$Waves}}\\	
  II(1) Start  		& 06:45:39  		& 75(F)-150(H) MHz 		& Artemis-IV, Hiras  \\
			& 			&   120 Km~sec$^{-1}$		& 	  \\
Release of Electron 	& 06:46-06:47 		& $\approx$127$-$225 keV	& \citet{Grechnev08} \\
 and Proton 		&  			& $\approx$1.7 Gev		&  \\
$\gamma$-ray peak	& 06:46-06:47		& $\approx$ 300 KeV-300 MeV	& SONG$/$CORONAS-F\\
 H$\alpha$ peak		& 06:46:30 		&	2B    			& LEAR \\
 II(1)	End		& 06:49 		& 60(F)-125(H) MHz			& Artemis-IV \\
 FCII(1) End		& 06:52 		& 120  MHz			& Artemis-IV \\
\hline
New set of Flare 	& 06:53-06:57		& 				& TRACE\\
Kernels activated	&       		& 				& \citep[cf.][]{Grechnev08}\\
CME First  		& 06:54 		& PA 293$^o$			& LASCO  \\
appearance on C2	&   			& 3242 Km~sec$^{-1}$, 		& \citet{Gopalswamy05} \\
			&   			& or 2075 Km$/$sec		& \citet{Grechnev08} \\
FCII(2) Start 		& 06:55			& 650 MHz, 			& Artemis-IV \\
        		&  			&  380 Km~sec$^{-1}$		&  \\
   HXR peak    		& 06:55 		&     				& RHESSI \\
  (secondary)  		&     			&                 		& \citep[][]{Grechnev08} \\
Electron release 	& 06:55 		&  				& \citet{Kurt10} \\
Microwave peak  	& 06:55 		& 9.4 GHz 			& NoRP \\
  (secondary)  		&     			&                 		&   \\
  II(2) Start  		& 06:56 		& 100(F)-200(H) MHz  		& Artemis-IV \\
			&			& 543 Km~sec$^{-1}$		&  \\
  III(2) GG      	& 06:57 		&				& Artemis-IV, Hiras\\
			&       		& 				& and {\em{Wind$/$Waves}}\\
II(2)	End		& 06:58 		& 75(F)-150(H) MHz		& Artemis-IV \\
 II(4) End  		& 07:02 		& 36 (F)-72 (H) MHz		& Artemis-IV \\
SXR Peak     		& 07:01-07:12 		& X7.1 plateau	 		& GOES 12\\
 FCII(2) End		& 07:03 		& 120  MHz			& Artemis-IV \\
\hline
 II(3)			& 07:15			& 14 MHz			& {\em{Wind$/$Waves}}\\
 			& 			& 750$-$4690 Km~sec$^{-1}$	& \citet{Pohjolainen07}\\
 SXR End       		& 07:26 		&				& GOES 12  \\
 H$\alpha$ End 		& 08:54 		&				& LEAR \\
\hline
\end{tabular}
\end{table}
%------------------------------------------------
\begin{figure}
\resizebox{\textwidth}{!}{\includegraphics{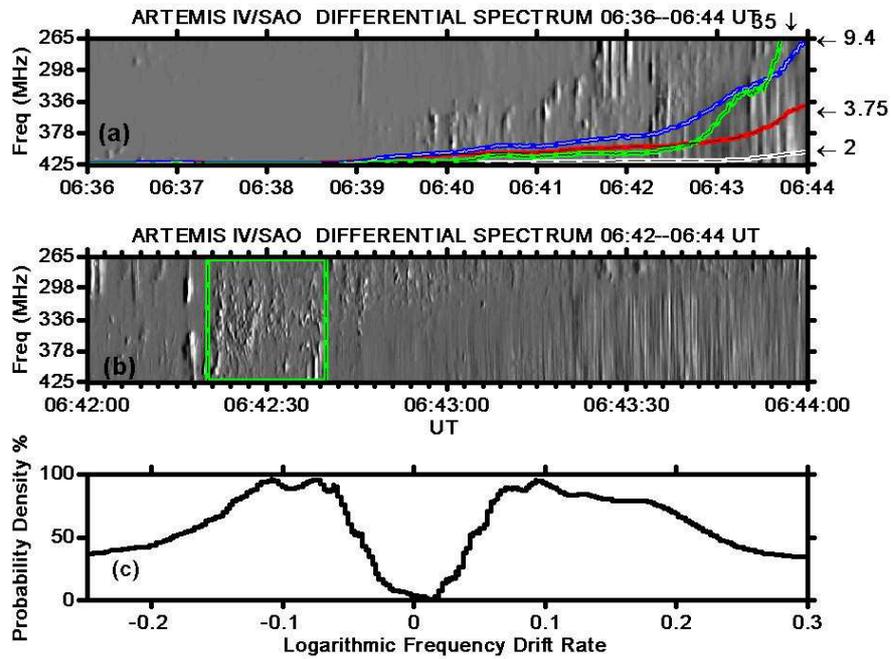}}
\caption{Narrow-band Type III and spikes high resolution (30 ms) dynamic spectrum
(a) Artemis-IV$/$SAO and the corresponding NoRPTeX
microwave enhancement (frequencies 35.0 (green), 9.4 (blue), 3.75 (red) and 2.2 (white) GHz).
This radio-activity in the 06:36\todash06:44 UT interval marks the onset of the
20 January 2005 event. (b) Artemis IV differential spectrum (SAO) in the 06:42\todash06:44 UT interval
with 10 msec resolution. On the left a group of spikes at 06:42:20 UT, marked by the frame,
on the right details of the pulsating {\em{patch}}.
(c) Evolution of average (logarithmic) frequency drift rate (df$/$fdt)of the marked spike cluster in the
period 06:42:20\todash06:42:40 UT; peaks appear at 0.10, $\pm$ 0.06, -0.11~se$c^{-1}$.
\citep[][also Appendix]{Tsitsipis06A,Tsitsipis07}.}
\label{fig02}
\end{figure}
%-------------------------------------------
% a landscape figure which scales and rotates the file f023.eps
\begin{sidewaysfigure}
\centering
\resizebox{\textwidth}{!}{\includegraphics{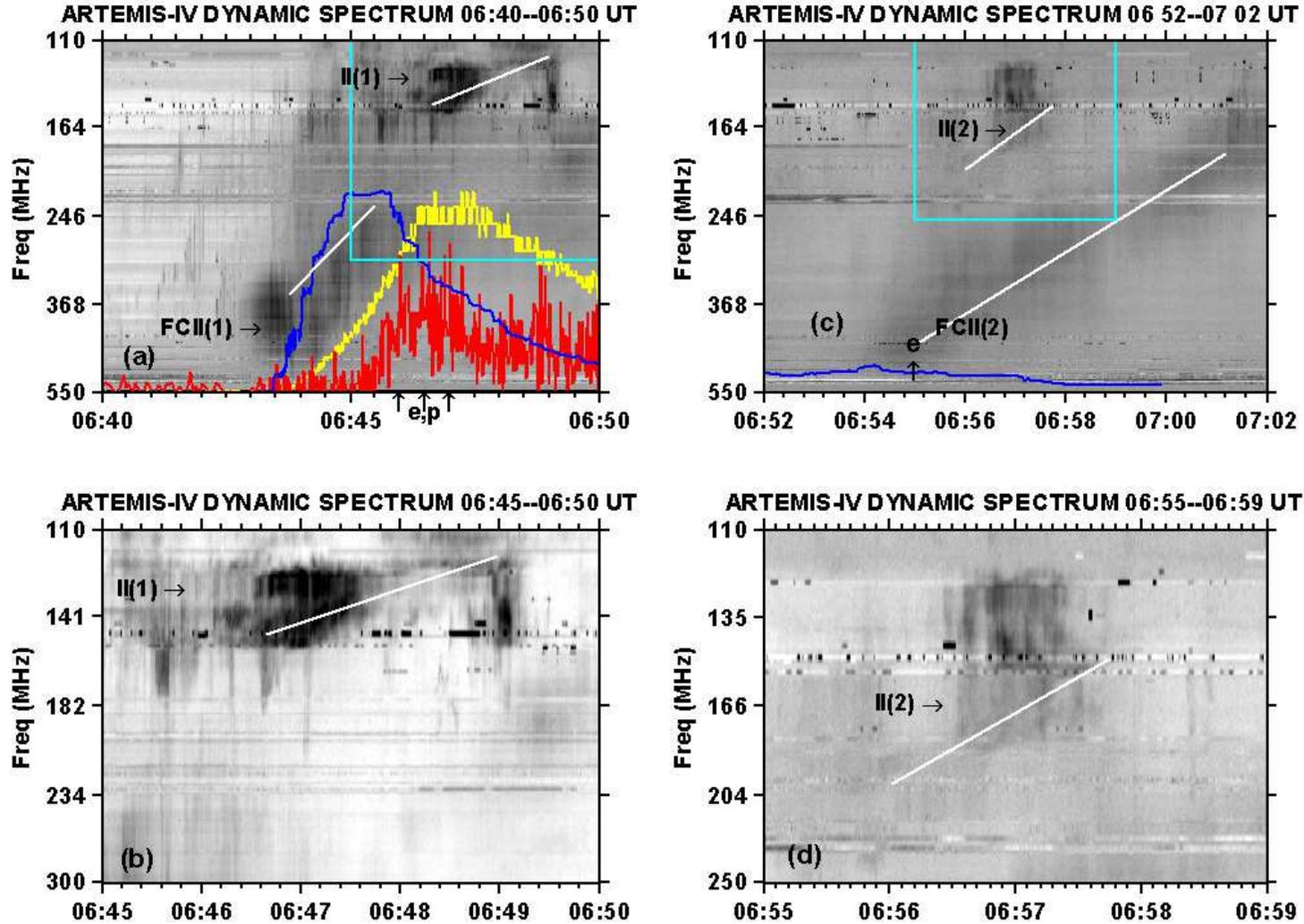}}
\caption{Artemis$-$IV (ASG) dynamic spectra of the Type II$/$FCII bursts
(see cyan frames on figure \ref{fig01}) (a) II(1)$/$FCII(1) 06:40-06:50 UT: 
The electron and proton release times (06:46-06:47 UT, Table \ref{T}) 
are marked with arrows under the plot. We have also included SONG$/$CORONAS-F normalized flux at 
40-100 KeV (blue), 0.775\todash2.0 MeV (yellow) and 60\todash100 Mev (red).
(b) Details of the Type II(1) shock in the 110-300 MHz range and 
the 06:45-06:50 UT time interval; it is enclosed by the box in (a).
(c) II(2)$/$FCII(2) 06:52-07:02 UT: The electron release time (06:55 UT, Table \ref{T}) 
is marked with an arrow. We have also included SONG$/$CORONAS-F normalized flux at 
40-100 Kev (blue). (d) Details of the Type II(2) shock in the 110-250 MHz range and 
the 06:55-06:59 UT time interval; it is enclosed by the box in (c).}
\label{fig01B}
\end{sidewaysfigure}
%-------------------------------------------
\begin{figure}
\resizebox{\textwidth}{!}{\includegraphics{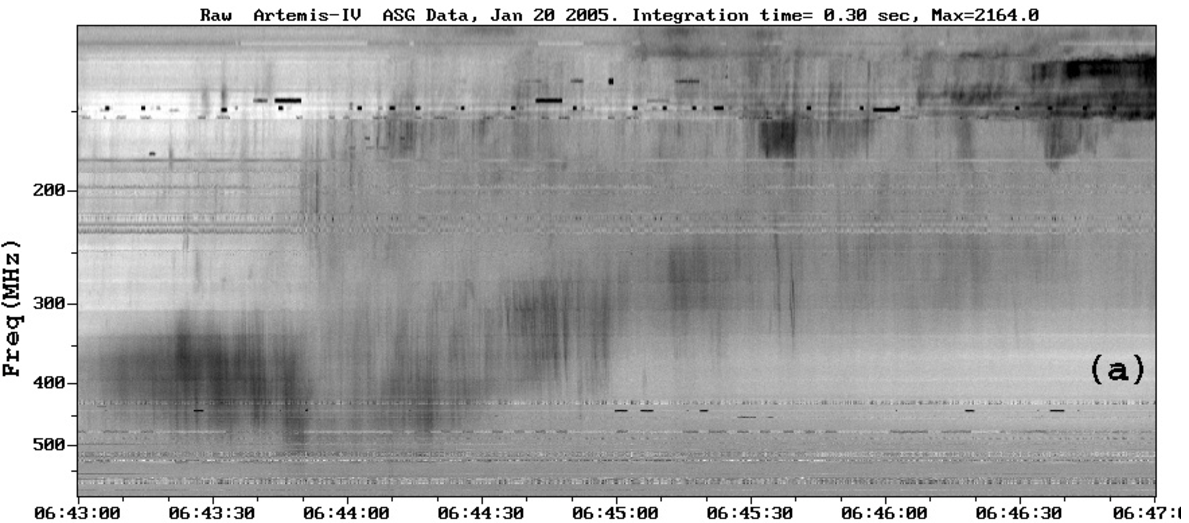}}
\resizebox{\textwidth}{!}{\includegraphics{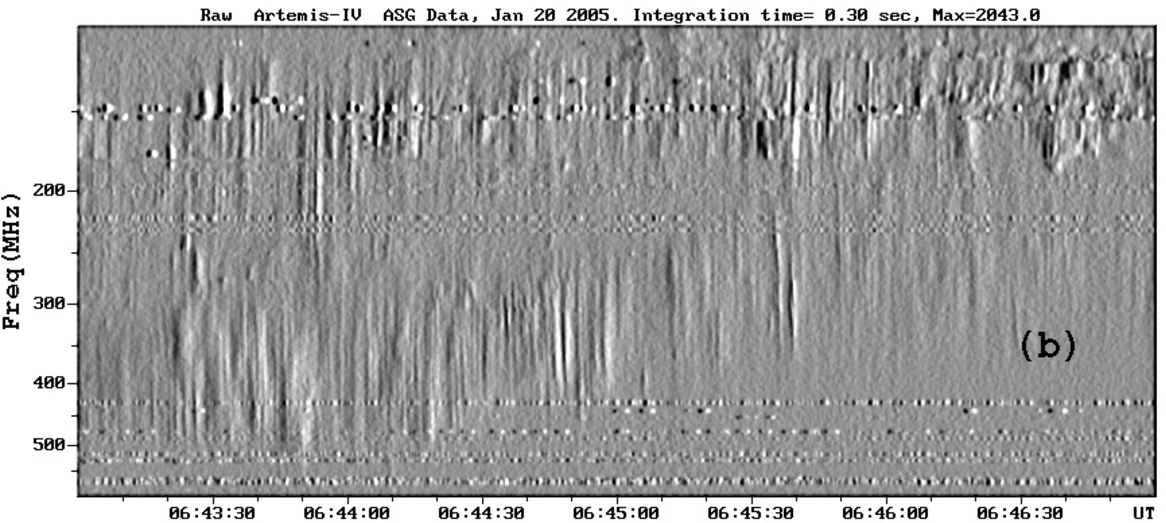}}
\resizebox{\textwidth}{!}{\includegraphics{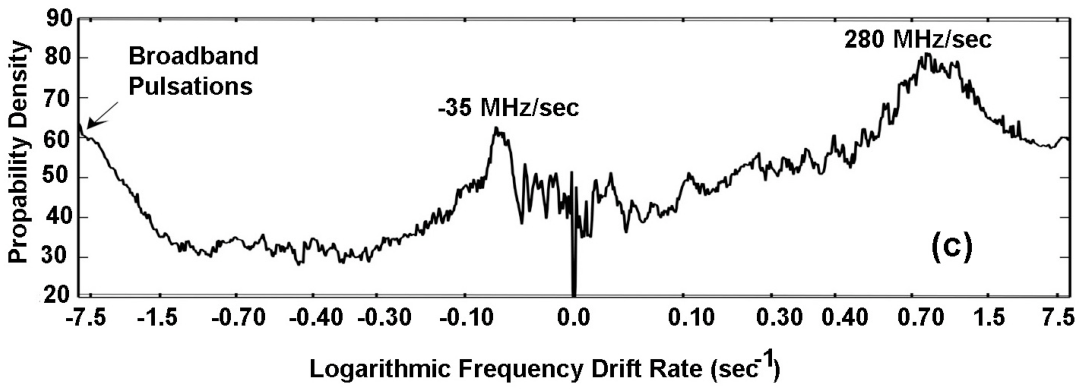}}
\resizebox{\textwidth}{!}{\includegraphics{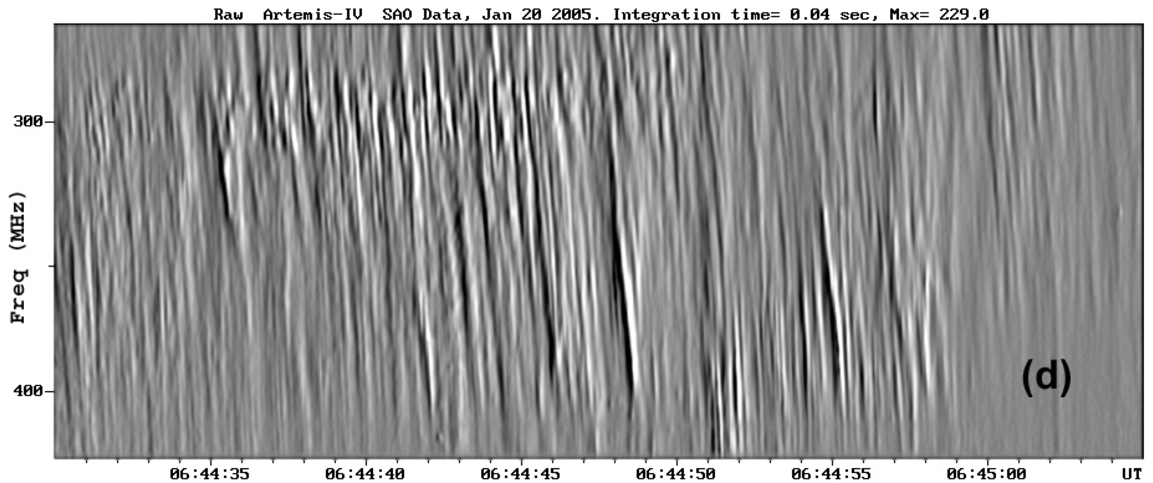}}
\caption{Artemis IV Spectra of II(1)$/$FCII(1), {\em{bidirectional}} Type III and reverse-drift
Type III-like bursts; (a) ASG dynamic spectrum (06:43-06:47 UT) (b) ASG differential spectrum,
(c) evolution of average (logarithmic) frequency drift rate (df$/$fd)
of the lowerpart of the differential spectrum (06:44-06:46 UT and 450-300 MHz)
(Appendix \ref{QL}) with peaks at -0.10~sec$^{-1}$~(20 Mm~sec$^{-1}$~outbound)
and 0.83~sec$^{-1}$~($\approx$ 120 Mm~sec$^{-1}$~reverse drift); we have assumed
radial propagation and a twofold Newkirk density-height coronal model
(d) SAO differential spectrum 06:44:30-06:45:05 UT.}
\label{fig04}
\end{figure}
%-------------------------------------------
\subsection{Coronal Density-Height Model Selection}
As plasma emission depends on electron density, which in turn may be converted to coronal height
using density models, we may calculate estimates for the radio source heights and
speeds from the dynamic spectra. The establishment of a 
frequency of observation-coronal height and frequency drift rate-radial speed correspondence
is affected by ambiguities introduced by the variation of the ambient medium properties.
These may be the result of burst exciter propagation within the undisturbed plasma, over-dense or
under-dense structure or CME after-flows
\citep[e.g.][for a detailed discussion on model selection]{Pohjolainen07,Pohjolainen08}. 

The twofold Newkirk model \citep{Newkirk} is adequate over active regions
at heights greater than 1.05 $R_{\odot}$ (frequencies lower than 300 MHz) \citep[figure 4 in ][]{Wild63},
and it is used throughout this report.

\section{Association of Active Phenomena with Radio Bursts}
\label{Event}
\subsection{Overview}
The complex radio bursts of the extreme event of 20 January 2005 have, attracted
considerable attention \citep [see e.g. ][ ]{Pohjolainen07, Masson09, Grechnev08, Mav09}.

The metric-radio-burst complex includes two distinct metric-decametric  
Type II (6:39-06:58 UT), in close succession, each associated in time with a 
flare-continuum-like burst (FCII), and Type III activity both isolated and in large groups (III GG).
These are superposed on an extended (06:36 UT to well past 08:00 UT) Type IV
continuum with rich fine structure (mostly fibers and pulsations). In the decametric
frequency range, a {stationary shock (with almost no drift) was appears in the 
40 (F) and 80 (H) MHz bands from 06:45:12 UT to 07:02 UT.

The radio bursts are 
accompanied by GOES$/$SXR and HXR, $\gamma$-ray emission in the 0.03-300 MeV 
range from SONG/CORONAS-F, and RHESSI; the high-energy channels, exhibit a sharp rise 
of intensity at 06:45:30 UT \citep[][]{Kuznetsov08,Grechnev08,Masson09}.

Small time-scale narrow-band features were recorded at high resolution 
by Artemis-IV$/$SAO, as part of the Type IV and FCII
fine structures. The latter provide evidence of magnetic reconnection
and electron acceleration in AR 720 (N14$^o$ W61$^o$).

An overview of the time sequence of the observations used in this report is presented in Table \ref{T}.
In Figure \ref{fig01} we present a combined dynamic spectrum of the radio bursts; we have included the
CME trajectory, the GOES SXR and SONG 40-100 KeV light-curve, 
and the times of energetic-particle injections for comparison. Details of the Type II events (II(1),
II(2) and II(4)) are presented in Figures \ref{Lin} and \ref{fig01B}.

In the following subsections, we have partitioned our description of the event, and the
corresponding radio spectra, for convenience. We present, in subsection \ref{onset}, the low-frequency part
of the dynamic spectrum as it applies to the hectometric Type II event (II(3)).
The description of the radio bursts during the early stages of
the flare is in subsection \ref{DM1}. In the next subsections, \ref{Complex01} and \ref{DM2}, we describe
the complex II-III GG-FCII bursts and the decametric stationary Type II in Section \ref{II4}. 
Lastly we present, briefly, the Type IV continuum in Subsection \ref{IV}.

\subsection{CME-flare Onset and DH Type II Shock} \label{onset}

The event of 20 January 2005 starts with the CME takeoff at 06:33 UT which accelerates
to high sky-plane speed at about 06:54 UT (3242 Km~sec$^{-1}$~after \citet{Gopalswamy05} or
2075 Km~sec$^{-1}$~from \citet{Grechnev08}). The GOES 12 SXR flare onset
at 06:36 UT reaches X7.1 intensity at 07:01-07:12 UT (Figure \ref{fig01});
its H$\alpha$ counterpart is a 2B flare from Learmonth (LEAR) in 06:41-08:54 with a peak at 06:46 UT. At,
06:36 UT the NoRP 3.75-35 GHz flux increase starts; this indicates that high-energy 
electrons were present from the very beginning of the event.

A hectometric  Type II event (II(3) in  Figure \ref{fig01}) appears at 07:15 UT at 14 MHz
on the {\em{Wind$/$Waves}}  spectra; \citet{Pohjolainen07} calculated the Type II(3) speed
from the {\em{Wind$/$Waves}} spectra to be in the 750-4690 Km~sec$^{-1}$~range (depending on the
selection of the coronal-density model).

\subsection{The Start of Decimetric-Metric Radio Activity}
\label{DM1}

The radio activity commences at the same time with the onset of the SXR flare
and the 3.75-35 GHz enhancement, at 06:36 UT (Figure \ref{fig02}), 
with a broad-band Type IV continuum which covers, almost,
the combined frequency range (2000-20 MHz) of Artemis-IV and Hiras (Figure \ref{fig01}); its duration
extends well past 08:00 UT (Section \ref{IV}).

A little later (06:39 UT), superposed on the continuum, groups of type J and U bursts at decimetric-metric
wavelengths appear. Their turnover frequencies drift slowly towards the lower part of the spectrum, 
forming a moving front (Figure \ref{fig02}a). The type J and U bursts
are accompanied by groups of narrow-band spikes. After the onset of the 100 keV HXR 
at 06:40 UT \citep{Masson09} the spike rate increases (06:42:20-06:42:40 UT, Figure \ref{fig02}b).
The spikes exhibit a frequency drift rate in the range -100 to 100 MHz~sec$^{-1}$~with distinct
peaks at {$\pm$0.06~sec$^{-1}$}, {$\pm$0.10~sec$^{-1}$} corresponding to exciter
speeds in the 10-16 Mm~sec$^{-1}$~range (Figure \ref{fig02}c).   The Type III(J) bursts
and the spikes, mentioned above, possibly mark an additional
acceleration episode corresponding to a small peak on the NoRP recordings.
At 06:43:00-06:43:45 UT a continuum patch 
appears on the dynamic spectra in the frequency range 1000-760 MHz (H) and 475-350 MHz (F)
(Figure \ref{fig01}). The continuum patch exhibits a pulsating structure (Figure \ref{fig02}) with periodic 
behaviour (0.32 Hz: main FFT peak, 10 Hz: secondary FFT peak) and coincides with the onset of the flare
continuum FCII(1). 

\subsubsection{The first Type II-FCII-III GG Complex Burst} \label{Complex01}

A close examination of FCII(1) (Figure \ref{fig04}) reveals extended groups of Type III
bursts and reverse-drift Type III bursts sharing a common onset frequency which corresponds to
the separation frequency of the bidirectional Type III bursts. The onset frequency drifts
towards lower frequencies. The high-frequency part of the FCII(1) (270-450 MHz) was recorded by 
the Artemis-IV$/$SAO with high time resolution ({100~sec$^{-1}$}). This part was examined
using the methodology introduced by \citet{Tsitsipis06A,Tsitsipis07}
which provides statistics on the frequency drift rate based on a 2D FFT of the dynamic spectrum
(Appendix \ref{QL}). The logarithmic frequency drift rates peak at -0.10 {sec$^{-1}$} and
0.17 {sec$^{-1}$} MHz~sec$^{-1}$~(speeds 15-20 Mm~sec$^{-1}$). 
A third and more pronounced peak at 0.83 {sec$^{-1}$} is characteristic 
of the reverse-drift Type IIIs. For the twofold Newkirk model the exciter speed is estimated 
to $\approx$ 120 Mm~sec$^{-1}$.

The Type II (II(1), in Figure \ref{fig01B}a and \ref{fig01B}b)
appeared on the Artemis-IV recordings at $\approx$ 06:45:40 UT and 145 MHz ending
at 06:49 UT and 125 MHz. The radial shock speed, as calculated from the twofold Newkirk coronal model, 
was found quite small ($\approx$ 120 Km~sec$^{-1}$).

The II(1), FCII(1), and III(1) combination of bursts appears to be associated with the rise phase
of the SXR flare (Figure \ref{fig01}). The intense Type III GG (III(1) in Figure \ref{fig01}) 
at 06:45:39 UT spans the whole range of the combined Hiras-Artemis IV-{\em{Wind$/$Waves}} spectra,
and it coincides in time with the particle release and the HXR-$\gamma$-microwave peaks
as specified in the following paragraph. 

The rise in microwave and 100 keV HXR flux (HXR enhancement starts at 06:40 UT, peak
at $\approx$ 06:45:30 UT) overlaps in time  with the drift of (FCII(1) $\approx$340 Km~sec$^{-1}$) to lower 
frequencies (6:44-6:46 UT) as it tends to converge with II(1), on the dynamic 
spectra, near 200 MHz ($\approx$ 0.26 {$R_{\odot}$}, Figures \ref{fig01} and \ref{fig01B}a).
This FCII(1)-II(1) convergence coincides with a sharp increase of $\gamma$-ray flux around 6:45-6:46 UT.
The latter originates from pion decay, as protons with energies in excess of 300 MeV
interact with  nuclei in the ambient corona, the pion decay time being $\approx$ 06:45 UT
\citep{Kuznetsov08,Grechnev08}. The pion decay emission supports the path and
time-of-flight calculations of \citet*{Grechnev08} which provide an estimate of the
proton release in the 06:46-06:47 UT interval. The electron release is estimated
within the same interval as the pion decay, and the HXR bremsstrahlung light curves
present similar profiles.

\subsubsection{The Second Type II-FCII-III GG Complex Burst} \label{DM2}

Another Type II occurence (II(2) in Figures \ref{fig01}, \ref{fig01B}c and \ref{fig01B}d starts 
at 06:56 and 200 MHz drifting at $\approx$ 543 km~sec$^{-1}$. It is accompanied by the FCII(2) 
event (starting 06:55 UT) which consists
mainly of a group of reverse-drift Type III events with a total drift
towards low frequencies corresponding to a radial exciter speed $\approx$ 380 Km~sec$^{-1}$.
It, also, coincides with a secondary maximum of the HXR \citep[][their Figure 5]{Grechnev08} 
associated with microwave flux enhancement, after 6:53 UT; to these
correspond an electron release at $\approx$06:55 UT and the second DH
Type III GG (III(2)) starting, on the dynamic spectrum, from the Type II(2) lane. 
This Type II event originated, probably, from a flare blast wave. 

The complex II(2)$/$FCII(2) burst is associated with the maximum of the SXR flux
but also with a weak HXR,  compared to the previous bursts, and a microwave peak.

\subsection{The Stationary Decametric Type II} \label{II4}
The Stationary Decametric Type II event 
starts at 06:45:16 UT at 41 MHz (Fundamental), exhibiting a fundamental-harmonic
structure; it is labeled II(4) on Figures \ref{fig01} and \ref{Lin}. 
The harmonic structure is not very well defined as it overlaps, somewhat,
with the fundamental of II(2) and it coincides, also, in frequency with the
Artemis-IV data gap. It ends at about 07:02 UT after the secondary 
energy release and the injection of the second Type III group (III(2)).
A number of Type III bursts appear to originate from the Type II band;
they merge into the two III GG groups (III(1) and III(2)).

\subsection{The Type IV Continuum} \label{IV}

The Type IV burst becomes more pronounced after II(2)$/$FCII(2);
it exhibits rich fine structure too which includes broad-band
pulsations, intermediate drift bursts (fibers) and a variety of narrow-band bursts.

Between FCII(2) and the stationary  Type IV a cluster of narrow-band structures, 
mostly Type III, was recorded by Artemis-IV/SAO 
at 300-400 MHz. A little after 07:00 UT, \citet{Grechnev08}
observed an expanding arcade, on GOES/SXI and EUV (SOHO/EIT) images, which may be
the source of the IV event.

\section{Summary and Discussion} \label{SandD}
We have presented a radio signature analysis of the solar eruptive event
of 20 January 2005 and we have studied the association of the radio bursts
with the energy release and the active
phenomena comprising this event.

\subsection{The  Dynamic Spectrum of 20 January 2005}\label{DynSpec}

{The two different Type II bursts (II(1) and II(2))  exhibited 
significantly different drift rates and were well separated  in time and
each was associated with an HXR peak. This suggests distinct shocks.}

{The first of them, II(1), was preceded by  metric-decimetric drifting structures
such as isolated Type III(J) and III(U) bursts, spikes and narrow-band Type III
events  (mostly reverse-slope); 
these can be identified as the signature of reconnection and electron acceleration 
episodes above expanding soft X-ray loops \citep{Klassen03}. 
Furthermore, the J-U bursts, at the beginning of the metric radio event,
trace the initial closed magnetic structure.}

Both, II(1) and II(2), are accompanied  by a flare continuum(FCII(1) and FCII(2) respectively)
which, on close examination, consists of bidirectional and reverse-drift Type III
elements; these continua also trace electrons accelerated in reconnection sites. 
The drift rate of the FCII suggests outward movement of the reconnection. 

The large Type III groups, III(1) and III(2), appear to start, in part,
from the Type II bands (II(1) and II(2) respectively) and in part from 
the decametric shock II(4) as discussed in Section \ref{II4} A few  
start at higher frequencies; this indicates
shock acceleration of a significant part of the electrons, exciting the Type III events.
Prior to the III GGs the Type III(J) or III(U) bursts trace
energetic electrons which are confined and do not access interplanetary space.
The continuation of III(1) and III(2) into hectometric frequencies
implies an opening of initially closed magnetic lines and
the escape of energetic particles into interplanetary space.
This agrees with \citet{Masson09}, as they report, before III(1), a
number of acceleration episodes in HXR of which the first few do not have DH radio signatures.
Within this context, the III(1) and III(2) onsets are consistent with the 
energetic-particle release times.

\subsection{Acceleration and Release of Energetic Particles within the Framework of the Standard Flare-CME model}

{The three main categories of particle acceleration processes 
\citep [e.g. ][ for a review]{Anastasiadis02,Aschwanden08} 
in solar flares and CMEs (electric DC-field acceleration at magnetic X-points,
stochastic acceleration by turbulence, and MHD shock acceleration in reconnection outflows)
are all present in the extreme event of 20 January 2005}. The DC-field acceleration 
appears in  reconnecting current sheet driven by the rising CME. The other two processes operate in the  
reconnection outflow termination shock accelerating
the highly energetic, electron and proton populations \citep{Mann06, Mann09, Warmuth09}.

%-------------------------------------------
\begin{figure}
\resizebox{\textwidth}{!}{\includegraphics{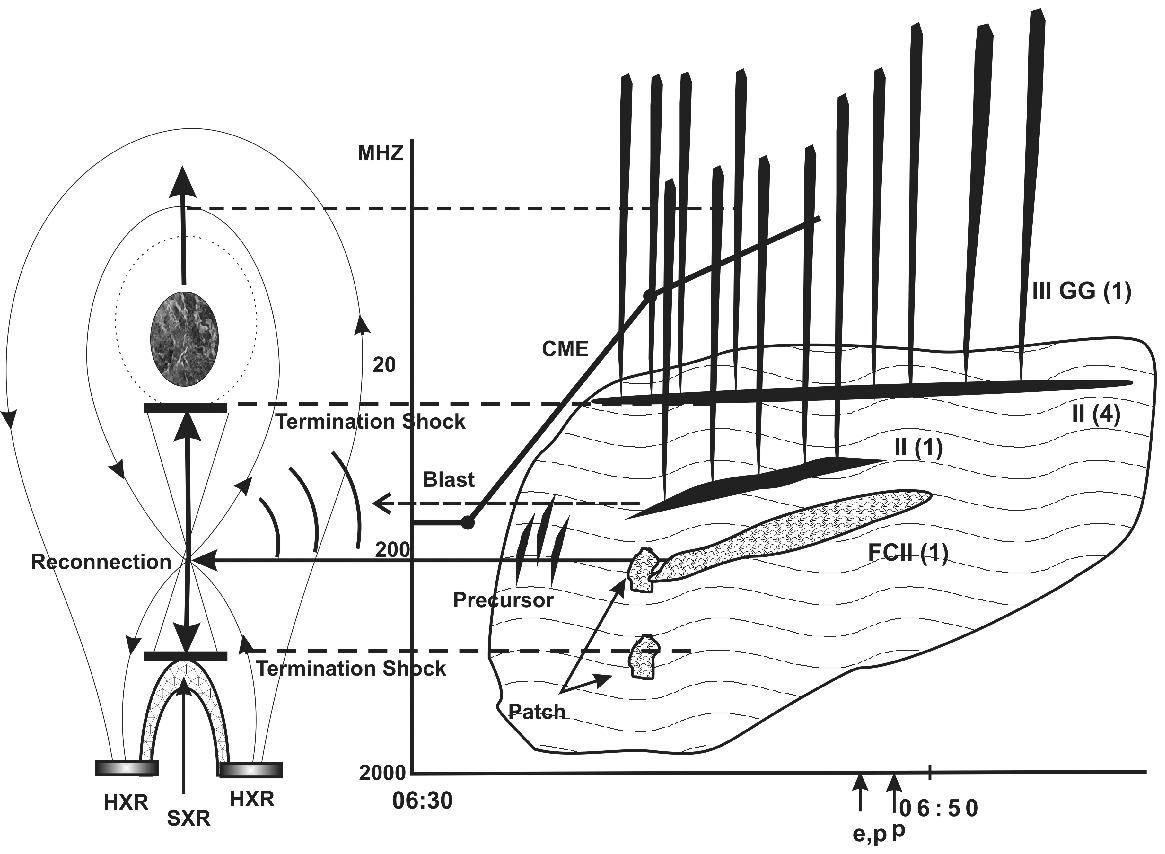}}
\caption{Comparison of the standard CME-Flare model with the
combined Hiras, Artemis-IV (ASG) and  {\em{Wind$/$Waves}} dynamic spectrum. 
On the left we present the CME-induced reconnection \citep{Forbes03}
supplemented with the reconnection outflow jets and the corresponding
termination shocks \citep{Aurass02} where the upward shock appears
after the CME. An additional Type II event originating from the flare-loop 
expansion moves sideways. On the right we present the dynamic spectrum 
resulting from this process.}
\label{FlCME}
\end{figure}
%-------------------------------------------
A close association, in time, of the electron injection at $\approx$06:45:40 UT, 
the HXR primary peak, and the start of the III GG burst (III(1)) was established
in Section \ref{Complex01}. The convergence of II(1)-FCII(1) spectra (06:45-06:47 UT), on the other
hand, coincides with the estimated time of proton escape at $\approx$06:45:30 \citep{Grechnev08}
as stated in Section \ref{DynSpec}. This temporal association seems to indicate that 
II(1) may be a termination shock formed as the reconnection front, FCII(1), rises; on the other hand
the expanding flare loops may {\em{piston drive}} II(1) as a flare blast \citep{Vrsnak08}. 
In both cases this shock formation may be accompanied by particle acceleration and release. 
What weighs more in favour of II(1) as a flare-associated Type II event is the appearance,
in the decametric frequency range, of II(4) whose spectrum appears to be more typical of 
a zero-drift termination shock; this also accelerates a significant part of the
Type III bursts in the groups III(1) and III(2).

In an attempt to introduce a model interpreting the observational data and the evolution of
the event, we firstly invoke the CME driven reconnection as proposed by
\citet{Forbes00}, \citet{Forbes03}, and \citet{Priest02}; the rising CME stretches and deforms
the coronal magnetic-field lines resulting in a vertical current sheet, where magnetic
reconnection commences. Since X-type reconnection is directly associated with
counterstreaming electron beams appearing as bidirectional Type IIIs
\citep{Aschwanden95,Aschwanden97} the FCII(1) is interpreted as a moving
reconnection front within the framework of this mechanism (Figure \ref{FlCME}). 

\citet{Kliem00} interpret pulsating structures such as the continuum patch marking the start of FCII(1) 
in terms of quasi-periodic particle acceleration episodes. These originate
from a magnetic reconnection in large-scale current sheets and are characterized by repeated
formation and coalescence of magnetic islands, feeding a continuously growing plasmoid
which drifts  outwards in the corona. In this case, however, there are no further observational
signatures of a rising plasmoid.

The slowly drifting burst II(4), may be a reconnection outflow termination shock 
as proposed by \citet{Aurass02} and \citet{Aurass03}. The only difference is that they
envisage a shock formed below the X-reconnection as the outflow encounters
the post-flare loop tops; our data, however, suggest that the shock is above the
reconnection as the Type II(4) is in the decametric range and always is at lower frequencies than the FCII(1).
Therefore we conclude that the shock may be formed from the upper outflow (Figure \ref{FlCME}).
The reconnection outflow termination shock was found to be capable of accelerating
the, highly energetic, electron and proton populations
indicated in the SONG recordings \citep{Mann06, Mann09, Warmuth09}. The presence of II(1)
complicates the spectral form somewhat; however it may originate at the flanks of the 
rising flare loops, hence it has a rather small speed.

The difficulties of the detection of the reconnection outflow termination shock on the
top of the post-flare loops are attributed by \citet{Aurass02} to the confusion on the dynamic
spectra due to other radio bursts, especially during the impulsive phase of the flare.

On the left of Figure \ref{FlCME} we propose a model including both
CME-induced reconnection \citep{Forbes03} and termination shocks from
reconnection outflow jets \citep{Aurass02};
the combined mechanisms may interpret the dynamic spectrum which we have
included on the right of the figure for comparison. At this point we note that
the II(1)$/$FCII(1)$/$III(1) complex burst was preceded by Type III(J) and
Type III(U) bursts, spikes, and the start of the Type IV continuum. This indicates
that energetic electrons were present before the onset of the main part of the
energy release in the flare, probably originating in the early stages of the
reconnection process.

During the second release of electrons, manifest as III(2) and associated with the secondary 
HXR peak reported by \citet{Grechnev08}, it is expected that the acceleration region has moved higher
as this HXR peak is quite small. The Type II burst (II(2)) drifts at $\approx$540 Km~sec$^{-1}$~and, being
far from the CME, may be interpreted as a flare blast at this point.
The whole Type II-flare continuum complex (II(2)-FCII(2)) follows the ignition of the, so far inactive, 
flare kernels (Section \ref{DM2}). \citet{Grechnev08} studied the TRACE~1600~\AA~images during the time
interval 6:52:41-6:57:30 UT, marked four pairs of kernels and computed intensity time profiles
of their brightness. As the brightness for two pairs of kernels decreases  it increases on the other two; 
this transition appears at about 6:53-6:54 UT. We note, at this point, that the decametric shock,
II(4), is still present until 07:02 UT, which is almost coincident with the end of the II(2)$/$FCII(2)
burst. This relatively long duration of the termination shock radio burst has already been reported,
by \citet{Aurass03b}.

\section{Conclusions} \label{Con}
The present study was based on high cadence data from the Artemis-IV solar radio-spectrograph
in the 650-20 MHz range, extended, as regards the frequency range, with Hiras (25-2500 MHz) 
and {\em{Wind$/$Waves}}
(13.825-1.075 MHz, 1040-20 kHz) recordings and supplemented with NoRP (2-35 GHz),
SONG (0.03-100 MeV), and GOES data; they were compared with the energetic-particle release times
reported in the literature. This made possible a  multi-frequency and multi-instrument
study of different aspects of the active phenomena, in the 20 January 2005 event.
Our analysis was focused on the {\em{Radio Perspective}} of the event, as radio observations provide
a rich collection of diverse diagnostics in the study of the shock-formation processes, the particle
acceleration, and the mass ejection from the Sun. The results were interpreted in terms of
the standard CME-flare model which applies to eruptive flares and invokes
the CME-induced reconnection scenario.

A number of particle accelerators were identified, and their relative importance in the manifestation
of the event was tested against their radio signatures. Firstly the CME-associated shock acceleration
seems to be of little importance. The reconnection in the wake of the CME and the reconnection outflow
termination shocks (both above and below the reconnection) appear to be the main contributors of
energetic electrons and protons.

The results are consistent with the standard CME-flare model with the {reconnection outflow
termination shocks and the reconnecting current sheet in the wake of the CME as sources of energetic particles.

%----------------------------------------------
\appendix   						% commands needed to redefine the labels belonging to the appendix
\renewcommand{\theequation}{A\arabic{equation}}    	% redefine the command that creates the equation no.
\setcounter{equation}{0}         			% reset equation counter
\renewcommand{\thefigure}{A\arabic{figure}}        	% redefine the command that creates the figure no.
\setcounter{figure}{0}           			% reset the counter value for the figures
\renewcommand{\thetable}{A\arabic{table}}           	% redefine the command that creates the table no.
\setcounter{table}{0}            			% reset the counter value for the tables
%----------------------------------------------
\section{Detection and Statistical Analysis of Quasi-Linear Structures embedded in Complex Dynamic Spectra}
\label{QL}
In the study of solar radio bursts, various types of structures often appear on the
dynamic spectra intertwined, overlapping, disturbed by terrestrial signals, and embedded
in a continuum background. A methodology \citep{Tsitsipis06A,Tsitsipis07} 
for the detection and statistical analysis
of linear structures has been developed and is presented here in brief.
The continuous background is eliminated by means of high pass filtering;
this suppresses slowly varying components enhancing fine structure with fast temporal
variation. The terrestrial interference, appears in the form of lines
parallel to the temporal axis and may be suppressed with directional filtering.

In the resulting clean dynamic spectrum, all frequency channels are normalized to
the same minimum and maximum; this eliminates variations of brightness along the frequency axis
which may interfere with the subsequent processing algorithms.

For the statistical analysis of quasi-linear structures representing Type III events,
fibers, pulsations etc., on the dynamic spectra we introduce the concept of energy density as a function of angle
which measures the signal power of a 2D image (a dynamic spectrum in this case) along a certain direction.
In terms of the two-dimensional Fourier transform of the dynamic spectrum:
\be
%------------------------------------------------------------
F\left( {\xi _1 ,\xi _2 } \right) = \int\limits_{ - \infty }^\infty  {\int\limits_{ - \infty }^\infty  {f\left( {x,y} \right) \cdot e^{ - j2\pi \left( {x\xi _1  + y\xi _2 } \right)} dxdy} }
%------------------------------------------------------------
\ee
The energy density as a function of angle (angular energy density) is defined as:
\be
%------------------------------------------------------------
S\left( \theta  \right) = 2\int\limits_0^\infty  {\left| {F\left( {\xi ,\theta } \right)} \right|} ^2 \xi d\xi 
%------------------------------------------------------------
\ee
This exhibits local maxima at the points where $\theta$ equals the
slopes of linear or quasi-linear segments within the image, yet it is not affected by their position within it.
These peaks of energy per direction on the image indicate, in a statistical sense, the dominant slopes of
linear or quasi-linear segments within the image, or, in this case, the peaks  of frequency drift rates
which may be interpreted as the dominant exciter velocities once a coronal density-height model has been
adopted. An example of the usage of this method appears in subsection \ref{DM1} and
in Figure \ref{fig04} (panels (c) and (e)), where we present the statistics of the frequency drift rates.

%----------------------------------------------
\begin{acks}
{We appreciate discussions and assistance of K-L Klein and
C. Caroubalos and C. Alissandrakis, the availability of the Nobeyama 
Radio Polarimetre data due to H. Nakajima and of the SONG
recordings due to B. Yu. Yushkov.
We also thank the anonymous referees for their comments and suggestions on the draft,
which have substantially improved the quality of this report.
This work was supported in part by the University of Athens Research Center (ELKE/EKPA).}
\end{acks}
%-------------------------

%-------------------------
}}
\end{article}
\end{document}